\magnification \magstep1
\raggedbottom
\openup 2\jot
\voffset6truemm
\def\cstok#1{\leavevmode\thinspace\hbox{\vrule\vtop{\vbox{\hrule\kern1pt
\hbox{\vphantom{\tt/}\thinspace{\tt#1}\thinspace}}
\kern1pt\hrule}\vrule}\thinspace}
\leftline {\bf Two-boundary problems in Euclidean quantum gravity}
\vskip 1cm
\leftline {GIAMPIERO ESPOSITO$^{1,2}$ and ALEXANDER
Yu. KAMENSHCHIK$^{3,4}$}
\vskip 1cm
\leftline {${ }^{1}$Istituto Nazionale di Fisica Nucleare,
Sezione di Napoli,}
\leftline {Complesso Universitario di Monte S. Angelo,}
\leftline {Via Cintia, Edificio G, 80126 Napoli, Italy}
\leftline {${ }^{2}$Universit\`a di Napoli
Federico II, Dipartimento di Scienze Fisiche,}
\leftline {Mostra d'Oltremare Padiglione 19, 80125 Napoli, Italy}
\leftline {${ }^{3}$L. D. Landau Institute for Theoretical
Physics of Russian Academy of Sciences,}
\leftline {Kosygina Str. 2, Moscow 117334, Russia}
\leftline {${ }^{4}$Landau Network-Centro Volta, Villa Olmo,
Via Cantoni 1, 22100 Como, Italy}
\vskip 1cm
\noindent
{\bf Summary}. - Recent work in the literature has studied
a new set of local boundary conditions for the quantized
gravitational field, where the spatial components of metric
perturbations, and ghost modes, are subject to Robin boundary
conditions, whereas normal components of metric perturbations
obey Dirichlet boundary conditions. Such boundary conditions
are here applied to evaluate the one-loop divergence on a
portion of flat Euclidean four-space bounded by two concentric
three-spheres.
\vskip 3cm
\leftline {PACS: 03.70, 04.60}
\vskip 100cm
\leftline {\bf 1. - Introduction}
\vskip 0.3cm
After several decades of work by many authors on the problems
of quantum gravity [1-11], 
it seems fair enough to say that the
path-integral approach remains an essential ingredient of any
attempt to understand the properties of the quantized gravitational
field. The crucial point is that quantum mechanics is a
physical theory whose predictions are of statistical nature.
When one tries to ``combine" it with general relativity, one
may thus expect to obtain a formalism where statistical concepts
as the partition function [12] find a natural place. This is
indeed the case for Euclidean field theories. This property is
possibly even more important than the opportunity to obtain a
space-time covariant approach to quantization, via the sum over
suitable classes of (or all) Riemannian four-geometries with their
topologies. Moreover, one knows that the effective action provides,
in principle, a tool for studying quantum theory as a theory of
small disturbances of the underlying classical theory, as well
as many non-perturbative properties in field theory [13-15].  

The basic object of a space-time covariant formulation of quantum
gravity may be viewed as being the path-integral representation 
of the $\langle {\rm out} | {\rm in} \rangle$ amplitude [13,14],
which involves the consideration of ghost fields that reflect the
gauge freedom of the classical theory [13,14]. In particular, what
seems to emerge is that the consideration of the elliptic
boundary-value problems of quantum gravity casts new light on 
the one-loop semiclassical approximation, which is the ``bridge" 
in between the classical world and the as yet unknown (full)
quantum theory [11]. We shall thus focus on this part of the
quantum gravity problem, i.e. the boundary conditions on metric
perturbations, when a Riemannian four-manifold (say $M$) with
boundary is considered (this may be a portion of flat Euclidean
four-space, or part of the de Sitter four-sphere, or a more
general curved background). To begin, we consider the problem of
imposing boundary conditions on the spatial components $h_{ij}$
of metric perturbations. Following ref. [16], we are interested
in Robin boundary conditions on $h_{ij}$. They are relevant for
the following reasons:
\vskip 0.3cm
\noindent
(i) They are part of a set of mixed boundary conditions of local
nature which ensure symmetry (and, with some care, self-adjointness)
of the elliptic operator acting on metric perturbations [17].
\vskip 0.3cm
\noindent
(ii) They admit, as a particular case, the boundary conditions on
the linearized magnetic curvature, which have a deep motivation in
several branches of classical and quantum gravity [18,19].
\vskip 0.3cm
\noindent
For simplicity, we study problems where the background
is totally flat, and curvature effects result from the boundary
only. All metric perturbations are then expanded on concentric
three-spheres of radius $\tau$, with $\tau \in [a,b]$, $a$ and
$b$ being the radii of the two bounding three-spheres. This
problem lies in between the quantum field-theoretical case, where
the boundary surfaces are by no means forced to be three-spheres, 
and may located in two asymptotic regions, and the quantum 
cosmological case, where one of the two boundary surfaces 
shrinks to a point [20]. The Robin-like boundary conditions 
proposed in ref. [16] read, therefore, 
$$
\left[{\partial h_{ij}\over \partial \tau}
+{\rho \over \tau}h_{ij}\right]_{\partial M}=0,
\eqno (1.1)
$$
where $\rho$ is a real parameter. Since an infinitesimal 
diffeomorphism changes $h_{ij}$ according to the law 
(hereafter, a vertical stroke denotes three-dimensional
covariant differentiation tangentially with respect to the
Levi--Civita connection of the boundary, and $K_{ij}$ is the
extrinsic-curvature tensor of the boundary)
$$
{ }^{\varphi}h_{ij}=h_{ij}+\varphi_{(i \mid j)}
+K_{ij}\varphi_{0} ,
\eqno (1.2)
$$
the request of being able to preserve (1.1) under the 
transformations (1.2) leads to the following boundary conditions
on normal and tangential components of the ghost 
one-form [16]:
$$
\left[{\partial \varphi_{0}\over \partial \tau}
+{(\rho+1)\over \tau}\varphi_{0}\right]_{\partial M}=0,
\eqno (1.3)
$$
$$
\left[{\partial \varphi_{i}\over \partial \tau}
+{\rho \over \tau}\varphi_{i}\right]_{\partial M}=0.
\eqno (1.4)
$$
The remaining boundary conditions are of the Dirichlet type
on normal components of metric perturbations:
$$
[h_{00}]_{\partial M}=0,
\eqno (1.5)
$$
$$
[h_{0i}]_{\partial M}=0.
\eqno (1.6)
$$
Regrettably, the invariance of {\it both} (1.5) and (1.6) under 
infinitesimal diffeomorphisms of metric perturbations is
incompatible with the boundary conditions (1.3) and (1.4), as was
proved in ref. [16]. Thus, we are studying a scheme where only
the $h_{ij}$ sector of the boundary conditions is gauge-invariant.
The expansions that we need are [21]
$$
h_{00}(x,\tau)=\sum_{n=1}^{\infty}a_{n}(\tau)Q^{(n)}(x),
\eqno (1.7)
$$
$$
h_{0i}(x,\tau)=\sum_{n=2}^{\infty}\left[b_{n}(\tau)
{Q_{\mid i}^{(n)}(x)\over (n^{2}-1)}
+c_{n}(\tau)S_{i}^{(n)}(x)\right],
\eqno (1.8)
$$
$$ \eqalignno{
\; & h_{ij}(x,\tau)=\sum_{n=3}^{\infty}u_{n}(\tau)
\left[Q_{\mid ij}^{(n)}(x)+{1\over 3}c_{ij}Q^{(n)}(x)
\right]+\sum_{n=1}^{\infty}{e_{n}(\tau)\over 3}
c_{ij}Q^{(n)}(x) \cr
&+\sum_{n=3}^{\infty}\left[f_{n}(\tau)
\Bigr(S_{i\mid j}^{(n)}(x)+S_{j \mid i}^{(n)}(x) \Bigr)
+z_{n}(\tau)G_{ij}^{(n)}(x)\right],
&(1.9)\cr}
$$
where $x$ are local coordinates on a three-sphere of radius
$\tau$. With a standard notation, $Q^{(n)}(x), S_{i}^{(n)}(x)$
and $G_{ij}^{(n)}(x)$ are scalar, transverse vector and 
transverse-traceless tensor harmonics on a unit three-sphere,
respectively [22]. 

Section {\bf 2} describes the way to implement the $\zeta$-function
method which is best suited for our analysis of one-loop
divergences. Sections {\bf 3}, {\bf 4}, 
{\bf 5} and {\bf 6} derive in detail the
contribution of transverse-traceless, vector, scalar and ghost
modes, respectively. 
Results and open problems are discussed in sect. {\bf 7}.
\vskip 5cm
\leftline {\bf 2. - $\zeta$-Function method}
\vskip 0.3cm
For a given elliptic operator, say $\cal A$, the spectral theorem
makes it possible to define its complex power ${\cal A}^{-s}$,
with $s \in C$ [11], and the $L^{2}$-trace of such a power is the
generalized $\zeta$-function for the operator $\cal A$:
$$
\zeta_{\cal A}(s) \equiv {\rm Tr}_{L^{2}}({\cal A}^{-s})
=\sum_{\lambda > 0} \lambda^{-s}.
\eqno (2.1)
$$
As is well known, the $\zeta$-function defined in (2.1) admits an
analytic continuation to the complex-$s$ plane as a meromorphic
function which is regular at $s=0$, so that the functional
determinant of the operator $\cal A$ may be {\it defined} by
the formula [23]
$$
{\rm det}({\cal A}) \equiv e^{-\zeta'(0)}.
\eqno (2.2)
$$
The value at the origin of the generalized $\zeta$-function 
contains all the information about the one-loop divergence and
the anomalous scaling factor of the amplitudes [11]. 

There exist, by now, several powerful algorithms for the evaluation
of $\zeta_{{\cal A}}(0)$. In particular, we are interested in the
technique developed in ref. [24] and applied several times since then
(see ref. [11] and references therein). Thus, we say that, denoting by
$f_{n}$ the function occurring in the equation obeyed by the 
eigenvalues by virtue of the boundary conditions, after taking
out false roots, and writing $d(n)$ for the degeneracy of the
eigenvalues parametrized by the integer $n$, one defines
the function
$$
I(M^{2},s) \equiv \sum_{n=n_{0}}^{\infty}d(n)n^{-2s}
\log f_{n}(M^{2}).
\eqno (2.3)
$$
What is crucial is the analytic continuation $``I(M^{2},s)"$
to the complex-$s$ plane of the function $I(M^{2},s)$, which
is a meromorphic function with a simple pole at $s=0$, i.e.
$$
``I(M^{2},s)"={I_{\rm pole}(M^{2})\over s}
+I^{R}(M^{2})+{\rm O}(s).
\eqno (2.4)
$$
The function $I_{\rm pole}$ is the residue at $s=0$, and makes it
possible to obtain the $\zeta(0)$ value as [11,24] 
$$
\zeta(0)=I_{\rm log}+I_{\rm pole}(M^{2}=\infty)
-I_{\rm pole}(M^{2}=0),
\eqno (2.5)
$$
where $I_{\rm log}$ is the coefficient of the $\log(M)$ term
in $I^{R}$ as $M \rightarrow \infty$. The contributions 
$I_{\rm log}$ and $I_{\rm pole}(\infty)$ are obtained from the
uniform asymptotic expansions of basis functions as 
$M \rightarrow \infty$ and their order $n \rightarrow \infty$,
while $I_{\rm pole}(0)$ is obtained by taking the
$M \rightarrow 0$ limit of the eigenvalue condition, and then
studying the asymptotics as $n \rightarrow \infty$. More 
precisely, $I_{\rm pole}(\infty)$ coincides with the coefficient
of ${1\over n}$ in the asymptotic expansion as 
$n \rightarrow \infty$ of
$$
{1\over 2}d(n)\log[\sigma_{\infty}(n)],
$$
where $\sigma_{\infty}(n)$ is the $n$-dependent term in the
eigenvalue condition as $M \rightarrow \infty$ and
$n \rightarrow \infty$. The $I_{\rm pole}(0)$ value is instead
obtained as the coefficient of ${1\over n}$ in the asymptotic 
expansion as $n \rightarrow \infty$ of
$$
{1\over 2}d(n)\log[\sigma_{0}(n)],
$$
where $\sigma_{0}(n)$ is the $n$-dependent term in the eigenvalue
condition as $M \rightarrow 0$ and $n \rightarrow 
\infty$ [11,24]. 
\vskip 0.3cm
\leftline {\bf 3. - Transverse-traceless modes}
\vskip 0.3cm
On using the de Donder gauge-averaging functional:
$$
\Phi_{a}(h) \equiv \nabla^{b}\Bigr(h_{ab}
-{1\over 2}g_{ab}g^{cd}h_{cd}\Bigr),
\eqno (3.1)
$$
the operator on metric perturbations reduces to the Laplacian
on symmetric rank-two tensors. Thus, the transverse-traceless
(TT) modes in the expansion (1.9) are found to take the 
form [21]
$$
z_{n}(\tau)=\alpha_{1} \tau I_{n}(M\tau)
+\alpha_{2} \tau K_{n}(M\tau),
\eqno (3.2)
$$
for all $n \geq 3$, the corresponding degeneracy being
$2(n^{2}-4)$. The boundary conditions (1.1) lead to
the equations
$$
\alpha_{1}\left(I_{n}'(M\tau_{+})+(\rho+1)
{I_{n}(M\tau_{+})\over M\tau_{+}}\right)
+\alpha_{2}\left(K_{n}'(M\tau_{+})+(\rho+1)
{K_{n}(M\tau_{+})\over M\tau_{+}}\right)=0,
\eqno (3.3)
$$
$$
\alpha_{1}\left(I_{n}'(M\tau_{-})+(\rho+1)
{I_{n}(M\tau_{-})\over M\tau_{-}}\right)
+\alpha_{2}\left(K_{n}'(M\tau_{-})+(\rho+1)
{K_{n}(M\tau_{-})\over M\tau_{-}}\right)=0.
\eqno (3.4)
$$
This implies that, to get rid of false roots, one has to
multiply by $M^{2}$ the resulting eigenvalue condition; on the
other hand, as $M \rightarrow \infty$, the eigenvalue condition
is proportional to $M^{-1}$. Thus, $I_{\rm log}$ is found to be
$$
I_{\rm log}={1\over 2}\sum_{n=3}^{\infty}2(n^{2}-4)(2-1)
=\zeta_{R}(-2)-4\zeta_{R}(0)+3=5.
\eqno (3.5)
$$
Moreover, as $n \rightarrow \infty$ and $M \rightarrow \infty$,
no $n$-dependent term occurs in the eigenvalue condition, which
implies
$$
I_{\rm pole}(\infty)=0.
\eqno (3.6)
$$
Last, as $M \rightarrow 0$ and $n \rightarrow \infty$, the
$\sigma_{0}(n)$ term in the eigenvalue condition reads 
$$
\sigma_{0}(n)=n \left(1-{(\rho+1)^{2}\over n^{2}}\right),
\eqno (3.7)
$$
which implies that no coefficient of ${1\over n}$ occurs in
the expansion of $(n^{2}-4)\log \sigma_{0}(n)$, and hence
$$
I_{\rm pole}(0)=0.
\eqno (3.8)
$$
The results (3.5), (3.6) and (3.8) imply that
$$
\zeta_{TT}(0)=5.
\eqno (3.9)
$$
Note that this contribution to $\zeta(0)$ has opposite sign,
with respect to the case when $h_{ij}$ perturbations are set
to zero at $\tau=\tau_{-}$ and $\tau=\tau_{+}$ [21]. 
\vskip 0.3cm
\leftline {\bf 4. - Vector modes}
\vskip 0.3cm
In the expansions (1.8) and (1.9) there is a decoupled vector
mode, $c_{2}(\tau)$, which reads [21]
$$
c_{2}(\tau)=\varepsilon I_{3}(M\tau)+\eta K_{3}(M\tau),
\eqno (4.1)
$$
and coupled vector modes, given by [21]
$$
c_{n}(\tau)={\widetilde \varepsilon}_{1}I_{n+1}(M\tau)
+{\widetilde \varepsilon}_{2}I_{n-1}(M\tau)
+\eta_{1}K_{n+1}(M\tau)+\eta_{2}K_{n-1}(M\tau),
\eqno (4.2)
$$
$$ \eqalignno{
f_{n}(\tau)&= \tau \biggr[-{1\over (n+2)}
{\widetilde \varepsilon}_{1}I_{n+1}(M\tau)
+{1\over (n-2)}{\widetilde \varepsilon}_{2}I_{n-1}(M\tau) \cr
&-{1\over (n+2)}\eta_{1}K_{n+1}(M\tau)
+{1\over (n-2)}\eta_{2}K_{n-1}(M\tau)\biggr],
&(4.3)\cr}
$$
with degeneracy $2(n^{2}-1)$. By virtue of (1.1) and (1.6), 
one has the boundary conditions 
$$
c_{2}(\tau_{+})=c_{2}(\tau_{-})=0,
\eqno (4.4)
$$
$$
c_{n}(\tau_{+})=c_{n}(\tau_{-})=0 \; \forall n \geq 3,
\eqno (4.5)
$$
$$
\left[{df_{n}\over d\tau}+{\rho \over \tau}f_{n}
\right]_{\tau=\tau_{+}}
=\left[{df_{n}\over d\tau}+{\rho \over \tau}f_{n}
\right]_{\tau=\tau_{-}}=0 \; \forall n \geq 3.
\eqno (4.6)
$$
As is well known, for decoupled (or finitely many) modes,
the contribution to $\zeta(0)$ is given by $I_{\rm log}$ only,
and for $c_{2}$ this reads 
$$
\zeta_{c_{2}}(0)={1\over 2}2(4-1)(0-1)=-3,
\eqno (4.7)
$$
because no false roots occur in the eigenvalue condition, whereas,
as $n \rightarrow \infty$ and $M \rightarrow \infty$, such 
eigenvalue condition is proportional to $M^{-1}$, picking up
a ${1\over \sqrt{M}}$ factor from both $I_{3}$ and $K_{3}$.

The eigenvalue condition resulting from the boundary conditions
(4.5) and (4.6) for coupled vector modes implies that one has
to multiply by $M^{2}$ to get rid of false roots. Moreover, as
$n \rightarrow \infty$ and $M \rightarrow \infty$, there is
proportionality to $M^{-2}$, and hence $I_{\rm log}$ is found
to vanish:
$$
I_{\rm log}={1\over 2}\sum_{n=3}^{\infty}2(n^{2}-1)(2-2)=0.
\eqno (4.8)
$$
To compute $I_{\rm pole}(\infty)$, one first evaluates
$\sigma_{\infty}(n)$, which is found to be
$$
\sigma_{\infty}(n)={4n^{2}\over (n^{2}-4)^{2}},
\eqno (4.9)
$$
and hence
$$
I_{\rm pole}(\infty)=0.
\eqno (4.10)
$$
Last, $\sigma_{0}(n)$ is found to be an even function of $n$
$$ \eqalignno{
\sigma_{0}(n)&={1\over (n^{2}-1)}\biggr[
{(2n^{4}-8n^{2}-2n^{2}\rho^{2}-4n^{2}\rho-8\rho^{2}-16\rho)
\over (n^{2}-4)^{2}} \cr
&+2{(n^{2}-\rho^{2}-2-2\rho)\over (n^{2}-4)}\biggr],
&(4.11)\cr}
$$
and hence $I_{\rm pole}(0)$ vanishes as well,
$$
I_{\rm pole}(0)=0,
\eqno (4.12)
$$
which implies
$$
\zeta_{c_{n},f_{n}}(0)=0.
\eqno (4.13)
$$
\vskip 0.3cm
\leftline {\bf 5. - Scalar modes}
\vskip 0.3cm
In the expansions (1.7)--(1.9), the scalar modes are 
$a_{n}(\tau),b_{n}(\tau),u_{n}(\tau)$ and $e_{n}(\tau)$. The modes 
$\left \{a_{1}(\tau),e_{1}(\tau) \right \}$, and
$\left \{a_{2}(\tau),b_{2}(\tau),e_{2}(\tau) \right \}$,
belong to finite-dimensional subspaces, and read [21] 
$$
a_{1}(\tau)={1\over \tau}\Bigr[\gamma_{1}I_{1}(M\tau)
+\gamma_{4}I_{3}(M\tau)+\delta_{1}K_{1}(M\tau)
+\delta_{4}K_{3}(M\tau)\Bigr],
\eqno (5.1)
$$
$$
e_{1}(\tau)=\tau \Bigr[3\gamma_{1}I_{1}(M\tau)
-\gamma_{4}I_{3}(M\tau)+3\delta_{1}K_{1}(M\tau)
-\delta_{4}K_{3}(M\tau)\Bigr],
\eqno (5.2)
$$
$$
a_{2}(\tau)={1\over \tau}\Bigr[\gamma_{1}I_{2}(M\tau)
+\gamma_{4}I_{4}(M\tau)+\delta_{1}K_{2}(M\tau)
+\delta_{4}K_{4}(M\tau)\Bigr],
\eqno (5.3)
$$
$$
b_{2}(\tau)=\gamma_{2}I_{2}(M\tau)-\gamma_{4}I_{4}(M\tau)
+\delta_{2}K_{2}(M\tau)-\delta_{4}K_{4}(M\tau),
\eqno (5.4)
$$
$$ \eqalignno{
e_{2}(\tau)&=\tau \Bigr[3\gamma_{1}I_{2}(M\tau)
-2\gamma_{2}I_{2}(M\tau)-\gamma_{4}I_{4}(M\tau) \cr
&+3\delta_{1}K_{2}(M\tau)-2\delta_{2}K_{2}(M\tau)
-\delta_{4}K_{4}(M\tau)\Bigr].
&(5.5)\cr}
$$
Moreover, for all $n \geq 3$, the scalar modes are all coupled,
and read [21] 
$$ \eqalignno{
a_{n}(\tau)&={1\over \tau}\Bigr[\gamma_{1}I_{n}(M\tau)
+\gamma_{3}I_{n-2}(M\tau)+\gamma_{4}I_{n+2}(M\tau) \cr
&+\delta_{1}K_{n}(M\tau)+\delta_{3}K_{n-2}(M\tau)
+\delta_{4}K_{n+2}(M\tau)\Bigr],
&(5.6)\cr}
$$
$$ \eqalignno{
b_{n}(\tau)&=\gamma_{2}I_{n}(M\tau)+(n+1)\gamma_{3}I_{n-2}(M\tau)
-(n-1)\gamma_{4}I_{n+2}(M\tau) \cr
&+\delta_{2}K_{n}(M\tau)+(n+1)\delta_{3}K_{n-2}(M\tau)
-(n-1)\delta_{4}K_{n+2}(M\tau),
&(5.7)\cr}
$$
$$ \eqalignno{
u_{n}(\tau)&=\tau \biggr[-\gamma_{2}I_{n}(M\tau)
+{(n+1)\over (n-2)}\gamma_{3}I_{n-2}(M\tau)
+{(n-1)\over (n+2)}\gamma_{4}I_{n+2}(M\tau) \cr
&-\delta_{2}K_{n}(M\tau)
+{(n+1)\over (n-2)}\delta_{3}K_{n-2}(M\tau)
+{(n-1)\over (n+2)}\delta_{4}K_{n+2}(M\tau)\biggr], 
&(5.8)\cr}
$$
$$ \eqalignno{
e_{n}(\tau)&=\tau \Bigr[3\gamma_{1}I_{n}(M\tau)
-2\gamma_{2}I_{n}(M\tau)-\gamma_{3}I_{n-2}(M\tau)
-\gamma_{4}I_{n+2}(M\tau) \cr
&+3\delta_{1}K_{n}(M\tau)-2\delta_{2}K_{n}(M\tau)
-\delta_{3}K_{n-2}(M\tau)-\delta_{4}K_{n+2}(M\tau)\Bigr],
&(5.9)\cr}
$$
with degeneracy $n^{2}$. Of course, it is the choice (3.1)
of gauge-averaging functional which leads to full agreement
with the formulae found in ref. [21] for the perturbative modes.

For the modes $a_{1}(\tau)$ and $e_{1}(\tau)$ the boundary 
conditions resulting from (1.5) and (1.1) are
$$
a_{1}(\tau_{+})=a_{1}(\tau_{-})=0,
\eqno (5.10)
$$
$$
\left[{de_{1}\over d\tau}+{\rho \over \tau}e_{1}
\right]_{\tau=\tau_{+}}
=\left[{de_{1}\over d\tau}+{\rho \over \tau}e_{1}
\right]_{\tau=\tau_{-}}=0.
\eqno (5.11)
$$
The Eqs. (5.10) and (5.11) lead to an eigenvalue condition
where one has to multiply by $M^{2}$ to get rid of false
roots. On the other hand, such eigenvalue condition is
proportional to $M^{-2}$ as $M \rightarrow \infty$. Thus, the
contribution to $\zeta(0)$ is found to vanish:
$$
\zeta_{a_{1},e_{1}}(0)={1\over 2} (2-2)=0.
\eqno (5.12)
$$

The modes $a_{2},b_{2},e_{2}$ obey, from sect. {\bf 1}, the boundary
conditions
$$
a_{2}(\tau_{+})=a_{2}(\tau_{-})=0,
\eqno (5.13)
$$
$$
b_{2}(\tau_{+})=b_{2}(\tau_{-})=0,
\eqno (5.14)
$$
$$
\left[{de_{2}\over d\tau}+{\rho \over \tau}e_{2}
\right]_{\tau=\tau_{+}}
=\left[{de_{2}\over d\tau}+{\rho \over \tau}e_{2}
\right]_{\tau=\tau_{-}}=0.
\eqno (5.15)
$$
In the resulting eigenvalue condition one has to multiply by
$M^{2}$ to get rid of false roots, whereas, as $M \rightarrow
\infty$, the eigenvalue condition is proportional to $M^{-3}$.
This property leads to a non-vanishing contribution to
$\zeta(0)$:
$$
\zeta_{a_{2},b_{2},e_{2}}(0)={1\over 2}\cdot 4 (2-3)=-2.
\eqno (5.16)
$$

Coupled scalar modes obey, for all $n \geq 3$, the boundary
conditions
$$
a_{n}(\tau_{+})=a_{n}(\tau_{-})=0,
\eqno (5.17)
$$
$$
b_{n}(\tau_{+})=b_{n}(\tau_{-})=0,
\eqno (5.18)
$$
$$
\left[{du_{n}\over d\tau}+{\rho \over \tau}u_{n}
\right]_{\tau=\tau_{+}}
=\left[{du_{n}\over d\tau}+{\rho \over \tau}u_{n}
\right]_{\tau=\tau_{-}}=0,
\eqno (5.19)
$$
$$
\left[{de_{n}\over d\tau}+{\rho \over \tau}e_{n}
\right]_{\tau=\tau_{+}}
=\left[{de_{n}\over d\tau}+{\rho \over \tau}e_{n}
\right]_{\tau=\tau_{-}}=0.
\eqno (5.20)
$$
The Eqs. (5.17)--(5.20) lead to an eigenvalue condition 
expressed by the vanishing of the determinant of an 
$8 \times 8$ matrix. However, the calculation is considerably
simplified if one remarks that, as $M \rightarrow \infty$, only
$K$ functions at $\tau=\tau_{-}$ and $I$ functions at 
$\tau=\tau_{+}$ give a non-negligible contribution [11,21]. 
Thus, the desired determinant splits into the product of two
determinants, say $D_{1}$ and $D_{2}$, of $4 \times 4$
matrices. As $M \rightarrow 0$, $D_{1}$ is proportional to
$M^{4n-2}$, and $D_{2}$ is proportional to $M^{-4n-2}$. Thus,
one has to multiply by $M^{4}$ the full determinant to get
rid of false roots. Moreover, both $D_{1}$ and $D_{2}$ are
proportional to $M^{-2}$ as $M \rightarrow \infty$, and hence
the full $I_{\rm log}$ vanishes:
$$
I_{\rm log}={1\over 2}\sum_{n=3}^{\infty}n^{2}(4-4)=0.
\eqno (5.21)
$$
To evaluate $I_{\rm pole}(\infty)$ and $I_{\rm pole}(0)$ we
note that, on defining
$$
\kappa \equiv \rho +1,
\eqno (5.22)
$$
$$
F_{+}(n) \equiv I_{n}'(M\tau_{+})
+\kappa {I_{n}(M\tau_{+})\over M\tau_{+}},
\eqno (5.23)
$$
$$
G_{-}(n) \equiv K_{n}'(M\tau_{-})
+\kappa {K_{n}(M\tau_{-})\over M\tau_{-}},
\eqno (5.24)
$$
one finds from (5.17)--(5.20) and (5.6)--(5.9) the 
fundamental formulae
$$ \eqalignno{
D_{1}(n)&=-{6n\over (n^{2}-4)}F_{+}(n-2)F_{+}(n+2)
I_{n}^{2}(M\tau_{+}) \cr
&-3{(n^{2}+1)\over (n+2)}F_{+}(n)F_{+}(n+2)I_{n}(M\tau_{+})
I_{n-2}(M\tau_{+}) \cr
&-3{(n^{2}+1)\over (n-2)}F_{+}(n)F_{+}(n-2)I_{n}(M\tau_{+})
I_{n+2}(M\tau_{+}) \cr
&-6n F_{+}^{2}(n)I_{n-2}(M\tau_{+})I_{n+2}(M\tau_{+}),
&(5.25)\cr}
$$
$$ \eqalignno{
D_{2}(n)&=-{6n\over (n^{2}-4)}G_{-}(n-2)G_{-}(n+2)
K_{n}^{2}(M\tau_{-}) \cr
&-3{(n^{2}+1)\over (n+2)}G_{-}(n)G_{-}(n+2)K_{n}(M\tau_{-})
K_{n-2}(M\tau_{-}) \cr
&-3{(n^{2}+1)\over (n-2)}G_{-}(n)G_{-}(n-2)K_{n}(M\tau_{-})
K_{n+2}(M\tau_{-}) \cr
&-6n G_{-}^{2}(n)K_{n-2}(M\tau_{-})K_{n+2}(M\tau_{-}).
&(5.26)\cr}
$$
By virtue of (5.25) and (5.26), the $n$-dependent term 
$D(n)=D_{1}(n)D_{2}(n)$ in the eigenvalue condition as
$M \rightarrow \infty$ and $n \rightarrow \infty$ is
$$
\sigma_{\infty}(n)={144 n^{2}(n^{2}-1)^{2}\over 
(n^{2}-4)^{2}}.
\eqno (5.27)
$$
This is an even function of $n$, and hence
$$
I_{\rm pole}(\infty)=0.
\eqno (5.28)
$$
Last, from the limiting form of modified Bessel functions as
$M \rightarrow 0$, one finds
$$
D_{1}(n)=-{3\Gamma^{-4}(n)(n-1)\over n^{3}(n+1)(n+2)(n^{2}-4)}
\Bigr[4n(n^{2}-1)(\kappa+n)^{2}-8(n^{2}+1)(\kappa+n)-8n \Bigr],
\eqno (5.29)
$$
$$
D_{2}(n)=-{3\Gamma^{4}(n)n(n+1)\over (n-1)(n-2)(n^{2}-4)}
\Bigr[4n(n^{2}-1)(\kappa-n)^{2}+8(n^{2}+1)(\kappa-n)-8n \Bigr].
\eqno (5.30)
$$
The results (5.29) and (5.30) lead to
$$
\sigma_{0}(n)=D(n)={9\over n^{2}(n^{2}-4)^{3}}H(n),
\eqno (5.31)
$$
where
$$ \eqalignno{
H(n) & \equiv 16n^{2}(n^{2}-1)^{2}(\kappa^{2}-n^{2})^{2}
+64 (n^{6}-n^{4}-3n^{2}-1)(\kappa^{2}-n^{2}) \cr
&-64n^{2}(n^{2}-1)(\kappa^{2}+n^{2})
+64n^{2}(2n^{2}+3),
&(5.32)\cr}
$$
which implies
$$
I_{\rm pole}(0)=0,
\eqno (5.33)
$$
because ${n^{2}\over 2}\log \sigma_{0}(n)$ is then an even
function of $n$. The Eqs. (5.21), (5.28) and (5.33) imply a
vanishing contribution to $\zeta(0)$,
$$
\zeta_{a_{n},b_{n},u_{n},e_{n}}(0)=0.
\eqno (5.34)
$$
\vskip 5cm
\leftline {\bf 6. - Ghost modes}
\vskip 0.3cm
The ghost one-form has a normal component, $\varphi_{0}(x,\tau)$,
and three tangential components, $\varphi_{i}(x,\tau)$. In our
problem, they are expanded on a family of concentric three-spheres
according to the relations [21] 
$$
\varphi_{0}(x,\tau)=\sum_{n=1}^{\infty} l_{n}(\tau)Q^{(n)}(x),
\eqno (6.1)
$$
$$
\varphi_{i}(x,\tau)=\sum_{n=2}^{\infty}\left[m_{n}(\tau)
{Q_{\mid i}^{(n)}(x)\over (n^{2}-1)}
+p_{n}(\tau)S_{i}^{(n)}(x)\right].
\eqno (6.2)
$$
By virtue of (3.1), the ghost operator reduces, in flat space,
to $-g_{ab}\cstok{\ }$, and hence one finds [21] 
$$
l_{1}(\tau)={1\over \tau}\Bigr[\kappa_{1}I_{2}(M\tau)
+\theta_{1}K_{2}(M\tau)\Bigr],
\eqno (6.3)
$$
$$
l_{n}(\tau)={1\over \tau}\Bigr[\kappa_{1}I_{n+1}(M\tau)
+\kappa_{2}I_{n-1}(M\tau)+\theta_{1} K_{n+1}(M\tau)
+\theta_{2}K_{n-1}(M\tau)\Bigr],
\eqno (6.4)
$$
$$ \eqalignno{
m_{n}(\tau)&=-(n-1)\kappa_{1}I_{n+1}(M\tau)+(n+1)\kappa_{2}
I_{n-1}(M\tau) \cr
&-(n-1)\theta_{1}K_{n+1}(M\tau)+(n+1)\theta_{2}K_{n-1}(M\tau),
&(6.5)\cr}
$$
$$
p_{n}(\tau)=\theta_{1} I_{n}(M\tau)+\theta_{2}K_{n}(M\tau).
\eqno (6.6)
$$
By virtue of (1.3), the decoupled ghost mode $l_{1}(\tau)$ obeys
the boundary conditions
$$
\left[{dl_{1}\over d\tau}+{(\rho+1)\over \tau}l_{1}
\right]_{\tau=\tau_{+}}
=\left[{dl_{1}\over d\tau}+{(\rho+1)\over \tau}l_{1}
\right]_{\tau=\tau_{-}}=0.
\eqno (6.7)
$$
The resulting eigenvalue condition is
$$
\left[{I_{2}'(M\tau_{+})\over M\tau_{+}}
+{\rho \over M^{2}\tau_{+}^{2}}I_{2}(M\tau_{+})\right]
\left[{K_{2}'(M\tau_{-})\over M\tau_{-}}
+{\rho \over M^{2}\tau_{-}^{2}}K_{2}(M\tau_{-})\right]=0.
\eqno (6.8)
$$
Hence one has to multiply by $M^{4}$ to get rid of false 
roots, whereas the behaviour of (6.8) as $M \rightarrow \infty$
is proportional to $M^{-3}$. This leads to
$$
\zeta_{l_{1}}(0)={1\over 2}(4-3)={1\over 2}.
\eqno (6.9)
$$

Coupled ghost modes are $l_{n}$ and $m_{n}$, for all $n \geq 2$.
In the light of (1.3) and (1.4), they obey the boundary conditions
$$
\left[{dl_{n}\over d\tau}+{(\rho+1)\over \tau}l_{n}
\right]_{\tau=\tau_{+}}
=\left[{dl_{n}\over d\tau}+{(\rho+1)\over \tau}l_{n}
\right]_{\tau=\tau_{-}}=0,
\eqno (6.10)
$$
$$
\left[{dm_{n}\over d\tau}+{\rho \over \tau}m_{n}
\right]_{\tau=\tau_{+}}
=\left[{dm_{n}\over d\tau}+{\rho \over \tau}m_{n}
\right]_{\tau=\tau_{-}}=0.
\eqno (6.11)
$$
In the resulting eigenvalue condition, one has to multiply 
by $M^{4}$ to get rid of false roots, whereas the behaviour
as $M \rightarrow \infty$ is proportional to $M^{-2}$, which
implies
$$
I_{\rm log}={1\over 2}\sum_{n=2}^{\infty}n^{2}(4-2)
=\zeta_{R}(-2)-1=-1.
\eqno (6.12)
$$
When $n \rightarrow \infty$ and $M \rightarrow \infty$, the
term $\sigma_{\infty}(n)$ in the eigenvalue condition is
$4n^{2}$, and hence
$$
I_{\rm pole}(\infty)=0.
\eqno (6.13)
$$
Moreover, when $M \rightarrow 0$ and $n \rightarrow \infty$,
the term $\sigma_{0}(n)$ in the eigenvalue condition is
$$
\sigma_{0}(n)={4n^{2}\over (n^{2}-1)}
\Bigr[(\rho^{2}+n^{2}-1)^{2}-4n^{2}\rho^{2}\Bigr].
\eqno (6.14)
$$
This is an even function of $n$, which implies
$$
I_{\rm pole}(0)=0,
\eqno (6.15)
$$
and, from (6.12), (6.13) and (6.15),
$$
\zeta_{l_{n},m_{n}}(0)=-1.
\eqno (6.16)
$$

Last, ghost vector modes obey, by virtue of (1.4), the
eigenvalue condition
$$
\left[I_{n}'(M\tau_{+})+{\rho \over M \tau_{+}}
I_{n}(M\tau_{+})\right]
\left[K_{n}'(M\tau_{-})+{\rho \over M \tau_{-}}
K_{n}(M\tau_{-})\right]=0.
\eqno (6.17)
$$
The resulting false roots are eliminated upon multiplication 
by $M^{2}$, whereas the behaviour of (6.17) as 
$n \rightarrow \infty$ and $M \rightarrow \infty$ is
proportional to $M^{-1}$, which implies
$$
I_{\rm log}={1\over 2}\sum_{n=2}^{\infty}2(n^{2}-1)(2-1)
=\zeta_{R}(-2)-\zeta_{R}(0)={1\over 2}.
\eqno (6.18)
$$
The Eq. (6.17) has no $n$-dependent term when $n \rightarrow 
\infty$ and $M \rightarrow \infty$, so that 
$$
I_{\rm pole}(\infty)=0.
\eqno (6.19)
$$
Last, as $M \rightarrow 0$ and $n \rightarrow \infty$, one finds
$$
\sigma_{0}(n)={1\over n}(\rho^{2}-n^{2}),
\eqno (6.20)
$$
and this leads to
$$
I_{\rm pole}(0)=0,
\eqno (6.21)
$$
$$
\zeta_{p_{n}}(0)={1\over 2}.
\eqno (6.22)
$$

The full $\zeta(0)$ value is the sum of the 9 contributions
given by Eqs. (3.9), (4.7), (4.13), (5.12), (5.16), (5.34),
(6.9), (6.16) and (6.22), i.e.
$$
\zeta(0)=5-3-2-2\left({1\over 2}-1+{1\over 2}\right)=0,
\eqno (6.23)
$$
where the round bracket is multiplied by $-2$ because ghost
fields for gravitation are fermionic and complex. Our result
agrees completely with the result expected for all 
two-boundary problems in the presence of a totally flat
Euclidean background (see the discussion in ref. [21]).
\vskip 5cm
\leftline {\bf 7. - Concluding remarks}
\vskip 0.3cm
The contribution of our paper is a detailed evaluation of the
one-loop divergence for the quantized gravitational field, by
studying all perturbative modes which contribute to the one-loop
Faddeev--Popov amplitude on a portion of flat Euclidean 
four-space bounded by two concentric three-spheres. The boundary
conditions used are (1.1) and (1.3)--(1.6), first proposed by
the authors in ref. [16]. Although a vanishing one-loop divergence
might have been expected on general ground, since the background is
totally flat, and only the gravitational field is considered, the
technical aspects of our analysis remain of some interest. As has
been shown in refs. [25,26], completely gauge-invariant boundary
conditions in Euclidean quantum gravity are in fact incompatible 
with the request of strong ellipticity of the boundary-value
problem. This is a technical condition, which amounts to requiring
that a unique solution should exist of the eigenvalue equation
for the leading symbol of the operator of Laplace type on metric
perturbations, subject to the boundary conditions and to an
asymptotic condition [25,26]. If this uniqueness fails to hold,
it is no longer possible to have a well defined form of one-loop
divergences, because the heat-kernel diagonal acquires a part which
is not integrable near the boundary [26].

Thus, the consideration of boundary conditions which are not
completely invariant under infinitesimal diffeomorphisms on metric
perturbations acquires new interest, since the lack of tangential
derivatives in the boundary operator makes it then possible to
satisfy the condition of strong ellipticity of the boundary-value
problem [25,26]. As far as we can see, at least three outstanding
problems should be now considered:
\vskip 0.3cm
\noindent
(i) Local boundary conditions along the lines of (1.1) and
(1.3)--(1.6) for curved backgrounds, with one or two boundary
surfaces.
\vskip 0.3cm
\noindent
(ii) The effect of the Prentki gauge for gravitation on manifolds
with boundary [27]. The resulting operator on metric perturbations
is no longer of Laplace type, and the corresponding form of
heat-kernel asymptotics on manifolds with boundary is largely
unexplored.
\vskip 0.3cm
\noindent
(iii) Inclusion of boundary operators of the integro-differential
type. For example, non-local boundary conditions for the Laplace
operator have been studied within the framework of Bose--Einstein
condensation models [28]. The counterpart for the gravitational 
field remains unknown, but could be studied by using the powerful
tools of functional calculus for pseudo-differential boundary
problems [29].
\vskip 0.3cm
\centerline {${ }^{*} \; \; { }^{*} \; \; { }^{*}$}
\vskip 0.3cm
A. KAMENSHCHIK was partially supported by RFBR via Grant No. 
96-02-16220, and is grateful to CARIPLO Science Foundation
for financial support.
\vskip 0.3cm
\leftline {REFERENCES}
\vskip 0.3cm
\item {[1]}
DeWITT B. S., {\it Rev. Mod. Phys.}, {\bf 29} (1957) 377.
\item {[2]}
MISNER C. W., {\it Rev. Mod. Phys.}, {\bf 29} (1957) 497.
\item {[3]}
WHEELER J. A., {\it Geometrodynamics} (Academic Press,
New York) 1962.
\item {[4]}
FEYNMAN R. P., {\it Acta Phys. Polonica}, {\bf 24} (1963) 697.
\item {[5]}
DeWITT B. S., {\it Phys. Rev.}, {\bf 160} (1967) 1113.
\item {[6]}
DeWITT B. S., {\it Phys. Rev.}, {\bf 162} (1967) 1195.
\item {[7]}
HAWKING S. W., in {\it General Relativity, an Einstein Centenary
Survey}, eds. S. W. Hawking and W. Israel (Cambridge University
Press, Cambridge) 1979.
\item {[8]}
ROVELLI C. and SMOLIN L., {\it Nucl. Phys. B}, {\bf 331}
(1990) 80.
\item {[9]}
ASHTEKAR A., {\it Lectures on Non-Perturbative Canonical 
Gravity} (World Scientific, Singapore) 1991.
\item {[10]}
GIBBONS G. W. and HAWKING S. W., {\it Euclidean Quantum Gravity}
(World Scientific, Singapore) 1993. 
\item {[11]}
ESPOSITO G., KAMENSHCHIK A. Yu. and POLLIFRONE G., {\it Euclidean
Quantum Gravity on Manifolds with Boundary}, 
in {\it Fundamental Theories
of Physics}, Vol. {\bf 85} (Kluwer, Dordrecht) 1997.
\item {[12]}
GIBBONS G. W. and HAWKING S. W., {\it Phys. Rev. D}, {\bf 15}
(1977) 2752.
\item {[13]}
DeWITT B. S., {\it Dynamical Theory of Groups and Fields}
(Gordon and Breach, New York) 1965.
\item {[14]}
DeWITT B. S., in {\it Relativity, Groups and Topology II}, eds.
B. S. DeWitt and R. Stora (North-Holland, Amsterdam) 1984.
\item {[15]}
JONA-LASINIO G., {\it Nuovo Cimento}, {\bf 34} (1964) 1790.
\item {[16]}
ESPOSITO G. and KAMENSHCHIK A. Yu., {\it Class. Quantum Grav.},
{\bf 12} (1995) 2715.
\item {[17]}
AVRAMIDI I. G., ESPOSITO G. and KAMENSHCHIK A. Yu., 
{\it Class. Quantum Grav.}, {\bf 13} (1996) 2361.
\item {[18]}
HAWKING S. W., {\it Phys. Lett. B}, {\bf 126} (1983) 175.
\item {[19]}
ESPOSITO G., {\it Quantum Gravity, Quantum Cosmology and
Lorentzian Geometries}, in {\it Lecture Notes in Physics, New Series
m: Monographs}, Vol. {\bf m12} (Springer-Verlag, Berlin) 1994.
\item {[20]} 
HARTLE J. B. and HAWKING S. W., {\it Phys. Rev. D},
{\bf 28} (1983) 2960.
\item {[21]}
ESPOSITO G., KAMENSHCHIK A. Yu., MISHAKOV I. V. and POLLIFRONE G.,
{\it Phys. Rev. D}, {\bf 50} (1994) 6329.
\item {[22]}
LIFSHITZ E. M. and KHALATNIKOV I. M., {\it Adv. Phys.},
{\bf 12} (1963) 185.
\item {[23]}
HAWKING S. W., {\it Commun. Math. Phys.}, {\bf 55} (1977) 133.
\item {[24]}
BARVINSKY A. O., KAMENSHCHIK A. Yu. and KARMAZIN I. P.,
{\it Ann. Phys. (N.Y.)}, {\bf 219} (1992) 201.
\item {[25]}
AVRAMIDI I. G. and ESPOSITO G., {\it Class. Quantum Grav.},
{\bf 15} (1998) 1141.
\item {[26]}
AVRAMIDI I. G. and ESPOSITO G., {\it Gauge Theories on
Manifolds with Boundary} (HEP-TH 9710048).
\item {[27]}
't HOOFT G. and VELTMAN G., {\it Ann. Inst. Henri Poincar\'e},
{\bf 20} (1974) 69.
\item {[28]}
SCHR\"{O}DER M., {\it Rep. Math. Phys.}, {\bf 27} (1989) 259.
\item {[29]}
GRUBB G., {\it Functional Calculus of Pseudodifferential 
Boundary Problems}, in {\it Progress in Mathematics}, Vol.
{\bf 65} (Birkh\"{a}user, Boston) 1996.

\bye